# NUCLEAR SPIN NOISE IN NMR REVISITED

Guillaume Ferrand,[1] Gaspard Huber,[2] Michel Luong,[1] Hervé Desvaux[2,a)]

1. Laboratoire d'ingénierie des systèmes accélérateurs et des hyperfréquences, SACM, CEA, Université Paris-Saclay, CEA/Saclay, F-91191 Gif-sur-Yvette, France.

2. Laboratoire Structure et Dynamique par Résonance Magnétique, NIMBE, CEA, CNRS, Université Paris-Saclay, CEA/Saclay, F-91191 Gif-sur-Yvette, France.

a) herve.desvaux@cea.fr

*Abstract*

*The theoretical shapes of nuclear spin-noise spectra in NMR are derived by considering a receiver circuit with finite preamplifier input impedance and a transmission line between the preamplifier and the probe. Using this model, it becomes possible to reproduce all observed experimental features: variation of the NMR resonance linewidth as a function of the transmission line phase, nuclear spin-noise signals appearing as a ``bump'' or as a ``dip'' superimposed on the average electronic noise level even for a spin system and probe at the same temperature, pure in-phase Lorentzian spin-noise signals exhibiting non-vanishing frequency shifts. Extensive comparisons to experimental measurements validate the model predictions, and define the conditions for obtaining pure in-phase Lorentzian-shape nuclear spin noise with a vanishing frequency shift, in other words, the conditions for simultaneously obtaining the Spin-Noise and Frequency-Shift Tuning Optima.*

## I. Introduction

Nuclear spin noise in NMR was initially predicted by Bloch in 1946 [1] and first detected at the end of the eighties by Sleator et al. using a SQUID for sensitive detection of the nuclear quadrupolar resonance of $^{35}$Cl at 4.2K [2, 3]. Later, McCoy and Ernst [4], and Leroy and Guéron [5], independently reported the observation of nuclear spin noise of protonated solvents using room temperature liquid-state NMR spectrometers. They took benefit of the narrow resonance lines of liquid-state NMR transitions and the strong coupling between large nuclear magnetization and the detection coil. There are two origins of nuclear spin noise in NMR. The first corresponds to the quantum fluctuation of the transverse magnetization and the second to the incoherent radiofrequency (RF) excitations of the longitudinal magnetization inducing in turn the appearance of transverse magnetization [3]. The RF excitations are produced by the Nyquist noise of the electronic detection circuit. Since the electronic circuit is resonant at the Larmor frequency, tiny magnetization fluctuations can be amplified when the nuclear magnetization is significantly coupled to the detection circuit. Consequently, the phenomenon of nuclear spin noise in NMR is strongly correlated to radiation damping [6, 7, 8], that is, the RF magnetic feedback field created by the current in the detection coil induced by the precessing transverse magnetization. This feedback field is expected to be in quadrature to the precessing magnetization for a perfectly tuned electronic circuit [9], inducing only broadening of the observed NMR signals for small flip-angle excitation pulses. Electronic mistuning lifts the perfect quadrature between the feedback field and the transverse magnetization whose NMR signatures are the previously mentioned signal broadening and a frequency shift of the observed NMR resonance frequency (frequency pushing) [10, 11].

After these early observations, nuclear spin noise has not attracted large attention from the NMR community, except for particular detection schemes dedicated to the monitoring of restricted numbers of spins [12, 13, 14]. Indeed, for usual NMR experiments with $10^{16}$ - $10^{20}$ spins, the coherent detection after RF excitation is several orders of magnitude more sensitive than noise detection scheme. Recently, the situation has changed [8, 15]. Detection or influence of nuclear spin noise were at the heart of studies of the initiation of spontaneous multiple maser emissions of hyperpolarized $^{129}$Xe [16, 17, 18], for the observation of other maser emissions [19, 20] or for the capability of continuous monitoring of hyperpolarized species without destroying the transient magnetization through RF excitations [21, 22]. Also, the appearance of cold-probes with very large quality factors $Q$ has facilitated the detection of nuclear spin noise. It has, for instance, allowed the direct acquisition of images without RF excitations [23] or the detection of thermally polarized $^{13}$C nuclear spin noise [24]; even 2D-NMR spectra based on spin-noise detection scheme have been reported [25]. Nevertheless, with consequences on a much broader audience these fundamental studies are mainly put forward problems of tuning of the electronic circuit [26]. Essentially when the electronic circuit is tuned at the Larmor frequency and matched at 50 Ω with respect to the emission circuit, a condition known as Conventional Tuning Optimum, CTO [27], the shapes of the nuclear spin-noise signals usually appear as distorted Lorentzian, contrary to the theoretical predictions [4, 10], revealing a mistuning according to the reception circuit. Conversely, when the electronic circuit was tuned for observing pure in-phase Lorentzian shapes for the nuclear spin-noise signals, an increase of the detected signals in conventional pulsed experiments was obtained with potentially an increase of signal-to-noise ratios [26]. This tuning condition was named Spin-Noise Tuning Optimum, SNTO [27]. More recently some of us have shown that even in conditions where spin-noise signals of pure in-phase Lorentzian shapes are observed, non-vanishing frequency pushing effects can be detected [28]. The tuning conditions, for which this frequency shift vanishes, have been named Frequency Shift Tuning Optimum. It was also demonstrated [28] that these experimental observations of the difference between FSTO and SNTO were not compatible with the theoretical predictions of references [4, 10].

The present article is devoted to solving this contradiction. We consider the complete detection circuit with a finite input impedance of the preamplifier and the effect of the transmission line length between the probe and the preamplifier as in Ref. [29]. We, in particular, take into account the noise produced by the preamplifier impedance which is fed back to the coil through the transmission line. The combination of the two last terms plays a key role for explaining the experimental observation. After Section II, dedicated to Materials and Methods, in Section III, theoretical models are described, providing evidence of the importance of both electronic elements. After a short description of the classical theory (Section III.A), in Section III.B it is predicted that, depending on the transmission line and preamplifier impedance, the radiation damping contribution can strongly vary altering the nuclear spin resonance line-width and potentially inducing, for a perfectly tuned system (SNTO condition), the appearance of a "bump" rather than the usual "dip" for the nuclear spin-noise signal superimposed on the average electronic noise level for a thermally equilibrated spin system and a classical probe. In addition in Section III.C the RF excitations affecting the nuclear susceptibility induced by Nyquist noise due to the preamplifier impedance are considered: a model allowing the numerical calculation of the nuclear spin-noise spectra is developed. Conversely to the model of Section III.B, using this model FSTO and SNTO conditions are not always simultaneously fulfilled. Finally, in Section III.D a mathematical framework is introduced which allows us to formally obtain the general shape of the nuclear spin-noise resonance. In this framework, it becomes possible to demonstrate that FSTO and SNTO conditions can differ, such an effect results from magnetization excitations induced by noise fluctuations within the preamplifier impedance combined with dephasing effects due to the transmission line. Section IV is devoted to the careful validation of the model: nuclear spin-noise spectra have been simulated using the measured electronic components and directly confronted to experimental measurements. In Section V, several aspects of the derivation are discussed; in particular an equation which can be used for determining physically relevant parameters is introduced. Finally, conclusions are drawn in Section VI.

## II.  *Materials and methods*

NMR experiments have been performed on a Bruker Avance DRX500 spectrometer equipped with a 5-mm inverse broadband probe-head with z-gradient, operating at 500 MHz for proton. In order to freely modify the preamplifier-probehead coupling, a phase shifter (ARRA inc., model 2448A), previously calibrated by connection to a vector network analyzer, was introduced between the preamplifier and the cable to the probe. The phase shift reference $\phi_0$ has been arbitrarily chosen. The total phase from the probe to the preamplifier $\phi$ was defined by $\phi = \phi_0 + \psi$, where $\psi$ was the variable phase difference introduced by the phase shifter, which conveniently simulates a transmission line of variable length.

For the quantitative validation between the numerical simulations and experiments, spectra were acquired at 293 K on a solution composed of 90% acetone and 10% deuterated chloroform for magnetic field lock purposes. First using a network analyzer directly connected to the probe, the capacitors $C_t$ and $C_m$ were set to ensure a probe circuit resonance frequency equal to the Larmor frequency and a perfect matching to 50 Ω. If the tuning and matching capacitors $C_t$ and $C_m$ were adjusted with the Bruker's wobulation routine, the values would have depended on the phase shift, preventing detailed comparisons between experiments and numerical simulations. This new procedure corresponds to the Probe Impedance Matching Optimum (PIMO) [30], the capacitances were kept constant and only the phase shift $\psi$ between the preamplifier and the probe was changed by adjusting the phase shifter from 0 to 191.9°, by increments of 10.1°. Due to the impedance of the TR-switch, the obtained capacitor values did not correspond to the CTO condition as determined using the Bruker's wobulation routine. For each of the 20 $\psi$ values, a spin-noise spectrum and a classical spectrum acquired thanks to a small-excitation pulse were acquired, and the measurements were compared to the values derived from numerical simulations performed with SciLab. All parameters used in these simulations were the experimental ones: the different unknowns (inductance, resistance of coil, $C_t$, $C_m$, input impedance of the preamplifier) have, in particular, been determined from external measurements performed with the network analyzer.

Spin-noise spectra were composed of 12 FIDs of 512k points, each acquired in 52.4 s. The latter were post-processed using SciLab with the sliding windows protocol for ensuring a final real resolution of about 0.7 Hz with a zero-fill factor of 2 and the optimal signal-to-noise ratio in the given amount of time [8, 21]. To avoid artifacts due to non-perfect SNTO conditions (leading to a mixture of absorptive and dispersive Lorentzian in spin-noise line-shapes), the resonance frequency for extracting frequency shifts and the line-widths of $^{12}CH_3COCH_3$ signal were determined from the excitation pulse spectra.

# III. Spin-Noise Theory

## A. Classical theory of nuclear spin-noise line-shape

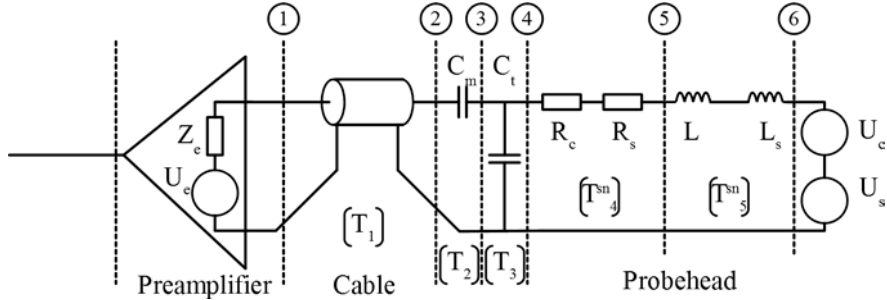

Figure 1: Schematic of the detection circuit with the different noise sources $U_e$, $U_c$ and $U_s$, for the preamplifier, coil and spin sources, respectively. The resistance $R_c$ and the inductance $L$ of the coil are affected by the presence of the nuclear susceptibility, represented by $R_s$ and $L_s$. The reference planes labeled with circled numbers allow the definition of the intermediate impedances and transmission matrices. $Z_{iL}$ and $Z_{iR}$ denote the impedances seen at plane $i$ from the preamplifier (left side) and from probe (right side), respectively.

Classically [9], the detection circuit is modeled by a coil ($L, R_c$) with a capacitor chip $C_t$ in parallel and the voltage is measured at the position of plane 3 of Figure 1. This measurement condition implicitly assumes a very high preamplifier input impedance, $Z_e$, directly connected at position of plane 3 in parallel to the capacitor chip $C_t$, i.e. without matching capacitor $C_m$ and transmission line. Following Mc Coy and Ernst [4], as a result of the spin resonance, the inductance of the coil becomes:

$$L \to L + L_s - \frac{jR_s}{\omega} = L(1 + \eta\chi). \tag{3.1}$$

The parameter $\eta$ corresponds to the filling factor, and $\chi = \chi' - j\chi''$ is the nuclear susceptibility:

$$\begin{cases} \chi' = \frac{\mu_0}{2}\gamma\, \mathcal{M}_z\, d(\delta\omega), \\ \chi'' = \frac{\mu_0}{2}\gamma\, \mathcal{M}_z\, a(\delta\omega). \end{cases} \tag{3.2}$$

In equation (3.2), $\mu_0$ is the magnetic permeability of free space, $\mathcal{M}_z$, the longitudinal nuclear magnetization which can be enhanced relative to thermal equilibrium magnetization $\mathcal{M}_0$ by hyperpolarization techniques [21, 22] or saturated by RF irradiation [4], $\gamma$ the magnetogyric ratio and $a(\delta\omega)$ and $d(\delta\omega)$ the absorptive and dispersive resonance line shapes, respectively:

$$\begin{cases} a(\delta\omega) = \frac{\lambda_2}{\lambda_2^2 + (\omega - \omega_0)^2}, \\ d(\delta\omega) = \frac{\omega - \omega_0}{\lambda_2^2 + (\omega - \omega_0)^2}, \end{cases} \tag{3.3}$$

with $\lambda_2$ the transverse self-relaxation rate and $\omega_0$ the nuclear spin Larmor resonance frequency. Then, introducing the quality factor $Q = L\omega_0/R_c$, the radiation damping rate $\lambda_r$ for a given longitudinal magnetization $\mathcal{M}_z$ is given by [4]:

$$\lambda_r = \frac{\mu_0}{2} \gamma \eta Q \mathcal{M}_z = \frac{\eta Q \chi''}{a(\delta\omega)} = \frac{\eta Q \chi'}{d(\delta\omega)}. \tag{3.4}$$

The radiation damping rate at thermal equilibrium $\lambda_r^0$ is defined by equation (3.4) with $\mathcal{M}_z = \mathcal{M}_0$. Due to the presence of the nuclear magnetization, the resistance and inductance of the coil have been replaced in Figure 1 by an equivalent spin-noise resistance and inductance $R_c^{sn} = R_c(1 + \lambda_r a(\delta\omega))$ and $L_c^{sn} = L + R_c \lambda_r d(\delta\omega)/\omega$, respectively. Associated to these resistances, the noise voltage is defined by $U_{sn} = U_R + U_s$, where $U_s$ and $U_R$ are the noise voltages generated by the spins and the resistance of the coil, respectively. We denote $W_R$ and $W_s$ their associated spectral densities:

$$\begin{cases} W_R = \frac{2}{\pi} k_b T_p R_c, \\ W_s = \frac{2}{\pi} k_b T_p R_c \lambda_r^0 a(\delta\omega). \end{cases} \tag{3.5}$$

where $T_p$ is the probe and sample temperature, $k_b$, the Boltzmann constant, and where we have used the remark that the spin term is temperature independent since it only depends on the number density of spins [4].

The voltage measurement is performed at the tuning capacitor chip terminals. In these conditions, according to Figure 1, the total spectral density is:

$$W_t = \frac{1}{\omega^2 C_t^2 R_c^2} \cdot \frac{\frac{2}{\pi} k_b T_p R_c + \frac{2}{\pi} k_b T_p R_c \lambda_r^0 a(\delta\omega)}{[1 + \lambda_r a(\delta\omega)]^2 + \left[\frac{1}{R_c}(L\omega - (C_t \omega)^{-1}) + \lambda_r d(\delta\omega)\right]^2} \tag{3.6}$$

Equation (3.6) simplifies when the Larmor resonance frequency, $\omega_0$, matches the electronic circuit resonance frequency, i.e. $\omega_0 \sqrt{LC_t} = 1$:

$$W_t = \frac{2}{\pi} \cdot \frac{k_B T_p}{\omega^2 C_t^2 R_c} \cdot \left[1 - \frac{\lambda_r^2 + 2\lambda_r \lambda_2 - \lambda_r^0 \lambda_2}{(\omega - \omega_0)^2 + (\lambda_r + \lambda_2)^2}\right]. \tag{3.7}$$

This corresponds to the equation derived by McCoy and Ernst [4] which has also been extended to the case of cold probes [27] or of hyperpolarized species [21].

### B. Nuclear spin-noise line shape with a noiseless preamplifier

We consider now a more realistic detection scheme where the input impedance of the preamplifier, $Z_e$, is not infinite [31] (cf. Figure 1). This has two consequences:

- A transmission line between the preamplifier and the probe is added whose length (or the associated phase $\phi$) provides an additional degree of freedom but which may also modify the response of the circuit if the input impedances of the probe and preamplifier are not matched to the transmission line impedance $Z_0$ (cf. Figure 1).

- The resistance impedance of the preamplifier $\Re(Z_e)$ [32] induces a new Nyquist noise source. This source can excite the nuclear magnetization through the induced current fluctuations due to the preamplifier resistance but in a non-straightforward way since there is no special relation between the impedance of the preamplifier and that of the probe.

Assuming, in the present subsection, that the noise temperature of the preamplifier $T_e$ is close to zero, the latter contribution to spin noise becomes negligible.

In Figure 1, $Z_{3R} = 1/(G + jB)$ was defined as the impedance seen by the coil at plane 3. Substituting $X = G^2 + (C_t\omega + B)^2$, the spectral density is:

$$W_t = \frac{2}{\pi} \cdot \frac{k_B T_p}{R_c X} \cdot \frac{1 + \lambda_r^0 a(\delta\omega)}{\left[1 + \frac{G}{R_c X} + \lambda_r a(\delta\omega)\right]^2 + \left[\frac{L\omega}{R_c} - \frac{C_t\omega + B}{X R_c} + \lambda_r d(\delta\omega)\right]^2}. \tag{3.8}$$

If $Z_{6R}$, the impedance seen by the voltage source $U_{sn}$ at reference plan 6, is real, then the source is in resonance with the circuit and the NMR frequency shift vanishes. This corresponds to FSTO condition:

$$L\omega = \frac{C_t\omega + B}{X}. \tag{3.9}$$

In fact, the solution of equation (3.9) depends on the impedance of the preamplifier $Z_e$ and of the phase of the transmission line $\phi$. In other words, FSTO depends on the transmission line length. Defining an apparent radiation damping rate $\lambda_r'$:

$$\lambda_r' = \frac{\lambda_r}{1 + \frac{G}{R_c X}}. \tag{3.10}$$

the expression of the spectral density becomes:

$$W_t = \frac{2}{\pi} \cdot \frac{k_B T_p}{R_c X \left(1 + \frac{G}{R_c X}\right)^2} \cdot \left(1 - \frac{\lambda_r'^2 + 2\lambda_r'\lambda_2 - \lambda_r^0\lambda_2}{(\omega - \omega_0)^2 + (\lambda_2 + \lambda_r')^2}\right). \tag{3.11}$$

In the specific case where $G = B = 0$, i.e. the preamplifier has an infinite impedance such as an open-circuit, then $X$ and $\lambda_r'$ are equal to $C_t^2\omega^2$ and $\lambda_r$ respectively, and Equation (3.11) reduces to Equation (3.7).

Finally, the measurement is performed through the voltage, $U_z$, at the extremities of the preamplifier input impedance. The measured spectral density can be computed from the voltage spectral density $W_t$ through the transmission line of impedance $Z_0$ and phase $\phi$ [33, 34]:

$$W_z = W_t \left|\cos\phi + j\cos\phi \frac{1}{Z_{3R} C_m \omega} - j\frac{Z_0 \sin\phi}{Z_{3R}}\right|^2. \tag{3.12}$$

According to equation 3.12, $W_z$ and $W_t$ exhibit same resonance line shapes since $\left|\cos\phi + j\cos\phi \frac{1}{Z_{3R} C_m \omega} - j\frac{Z_0 \sin\phi}{Z_{3R}}\right|^2$ does not depend on the nuclear spin susceptibility and consequently

varies slowly compared to $W_t$ around the Larmor frequency $\omega_0$. Furthermore, according to equations (3.11), in the FSTO condition, under the assumption of vanishing noise temperature of the preamplifier, the shape of $W_t$, i.e. the spin-noise shape of $W_z$, is an in-phase Lorentzian and corresponds to the SNTO condition. Since FSTO and SNTO conditions always match, this model is too simplified for explaining the observed discrepancies reported in Ref. [28].

## C. Model with preamplifier noise

To take into account the effects of RF excitations induced by the finite impedance of the preamplifier on the nuclear spins and their consequences on spin-noise spectra ($T_e \neq 0$), we shall, in this section use a transmission matrix, also known as ABCD matrix, approach. The whole system is described by a single matrix $T^{sn} = T_1 T_2 T_3 T_4^{sn} T_5^{sn}$ of dimension 2×2. The transmission matrices $T_i$ are given in microwave engineering textbooks [34]:

$$\begin{cases} T_1 = \begin{pmatrix} \cos\phi & jZ_0 \sin\phi \\ \frac{j\sin\phi}{Z_0} & \cos\phi \end{pmatrix}, T_2 = \begin{pmatrix} 1 & \frac{1}{jC_m\omega} \\ 0 & 1 \end{pmatrix}, \\ T_3 = \begin{pmatrix} 1 & 0 \\ jC_t\omega & 1 \end{pmatrix}, T_4^{sn} = \begin{pmatrix} 1 & jL_c^{sn}\omega \\ 0 & 1 \end{pmatrix}, T_5^{sn} = \begin{pmatrix} 1 & R_c^{sn} \\ 0 & 1 \end{pmatrix}. \end{cases} \quad (3.13)$$

The measured voltage $U_z$ across the preamplifier impedance $Z_e$ is given by:

$$\begin{pmatrix} U_z \\ i_z \end{pmatrix} = \begin{pmatrix} T_{11}^{sn} & T_{12}^{sn} \\ T_{21}^{sn} & T_{22}^{sn} \end{pmatrix} \begin{pmatrix} U_{sn} \\ i_{sn} \end{pmatrix}, \quad (3.14)$$

where $T_{ij}^{sn}$ are elements of matrix $T^{sn}$. Introducing $Z_e$ that relates $U_z$ and $i_z$, one obtains [35]:

$$W_z = \left| \frac{Z_e}{Z_e T_{22}^{sn} + T_{12}^{sn}} \right|^2 W_{sn}, \quad (3.15)$$

with $W_{sn} = W_R + W_s$. In the FSTO condition, the result is given by equation (3.12).

Noise contribution from a preamplifier is classically modeled with real part of its input impedance combined to an equivalent temperature $T_e$ [34]; therefore, the preamplifier noise spectral density is:

$$W_{pa} = \frac{2}{\pi} kT_e \Re(Z_e). \quad (3.16)$$

The impedance seen by the preamplifier, $Z_{1L}$, is given by $Z_{1L} = T_{12}^{sn}/T_{22}^{sn}$, and the contribution of the preamplifier to the spin-noise spectrum $W_a$ by:

$$W_a = \left| \frac{T_{12}^{sn}}{Z_e T_{22}^{sn} + T_{12}^{sn}} \right|^2 \cdot \frac{2}{\pi} kT_e \Re(Z_e). \quad (3.17)$$

Comparison of Eqs. (3.15) and (3.17) indicates a different dependence on the nuclear susceptibility of these two noise contributions since $T_{12}^{sn}$ depends on $L_c^{sn}$ and $R_c^{sn}$. In addition to the probe, spins and preamplifier noise contributions to the measured spectra, there is an extra component which is constant over the frequency range and does not interact either with the probe or the spins: the constant noise of the preamplifier and the noise coming from other external sources between the preamplifier and the analog-to-digital converter (ADC), hereafter denoted $W^{ext}$. It introduces an offset in the spectra and cannot be neglected if comparisons to experimental spectra are carried out. Hence, the simulated spin-noise power spectrum should be written:

$$W_{tot} = \frac{2kT_p}{\pi}|T_{12}^{sn}|^2 \frac{(R_c + R^{sn}) + \frac{T_e}{T_p}\Re(Z_e)|T_{22}^{sn}|^2}{|Z_{1L}T_{22}^{sn} + T_{12}^{sn}|^2 \cdot |T_{22}^{sn}|^2} + W^{ext}. \quad (3.18)$$

### D. Pseudo-Lorentzian shape of the spin-noise spectrum

For extracting NMR relevant parameters without having access to all electronic parameters (the usual situation in particular with cold probes) we show here that the general shape of nuclear spin noise is a generalized Lorentzian function (called pseudo-Lorentzian function) [35]. The pseudo-Lorentzian function, $\Lambda[p,q](x)$ with $p$ and $q$ complex-valued parameters, is defined by:

$$\Lambda[p,q](x) = 1 + \frac{p}{1 + jqx} \quad (3.19)$$

This function has several properties [35] but an important one for the present purpose, is the capability to rewrite any pseudo-Lorentzian function with a real-value parameter $|q|^2/\Re(q)$ and a shift $(-\Im(q)/|q|^2)$ according to the $x$ parameter:

$$\text{For } q \neq 0, \quad \Lambda[p,q](x) = 1 + \frac{1}{\Re(q)}\frac{pq^\star}{1 + j\frac{|q|^2}{\Re(q)}\left(x - \frac{\Im(q)}{|q|^2}\right)}. \quad (3.20)$$

The total impedance of the coil with the nuclear spin contributions can therefore be split into two contributions: the coil inductance $jL\omega$ and the residual impedance of the coil, $Z_{rs}$, defined by:

$$Z_{rs} = R_c + jL\omega\eta(\chi' - j\chi''). \quad (3.21)$$

Using the definition of the pseudo-Lorentzian function (3.19), $Z_{rs}$ can be written as:

$$Z_{rs} = R_c \cdot \Lambda\left[\frac{\lambda_r}{\lambda_2}, -\frac{\omega_0}{\lambda_2}\right]\left(\frac{\omega - \omega_0}{\omega_0}\right). \quad (3.22)$$

We consider the transmission matrix $T^R$ with $T^R = T_1 T_2 T_3 T_L$, where $T_L$ is the transmission matrix of the coil inductance $L$. The impedance $Z_{1L}$ seen by the preamplifier is:

$$Z_{1L} = \frac{Z_{rs}T_{11}^R + T_{12}^R}{Z_{rs}T_{21}^R + T_{22}^R}. \quad (3.23)$$

After little algebra with transmission matrices and using several properties of the pseudo-Lorentzian function one can show that the impedance $Z_{1L}$ is a pseudo-Lorentzian function whatever the values of $T^R$. The Johnson-Nyquist noise, $W_z$, produced by this passive system is:

$$W_z = \frac{2}{\pi} k_b T_p \Re(Z_{1L}) \frac{|Z_e|^2}{|Z_{1L} + Z_e|^2}, \qquad (3.24)$$

Introducing the preamplifier noise source, $W_a$, and the external noise source, $W^{ext}$, as in equation (3.18), provides an equation for the measured spin noise:

$$W_{tot} = \frac{2}{\pi} k_B \cdot \left( \frac{T_e \Re(Z_e)|Z_{1L}|^2}{|Z_{1L} + Z_e|^2} + \frac{T_p \Re(Z_{1L})|Z_e|^2}{|Z_{1L} + Z_e|^2} \right) + W^{ext}. \qquad (3.25)$$

Using the properties of the real part, Equation (3.25) can be written as:

$$W_{tot} = \frac{2}{\pi} k_B \cdot T_e \cdot \Re\left( \frac{Z_e Z_{1L}}{Z_e + Z_{1L}} \cdot \left( \frac{\frac{T_p}{T_e} \cdot Z_e + Z_{1L}}{Z_e + Z_{1L}} \right)^* \right) + W^{ext}. \qquad (3.26)$$

Since it can be shown that the term $\frac{Z_e Z_{1L}}{Z_e + Z_{1L}}$ is a pseudo-Lorentzian function, we shall denote its associated $q$ parameter as $q_0$. Using properties of the pseudo-Lorentzian function, $W_{tot}$ can be written as [35]:

$$W_{tot} = \frac{2}{\pi} k_B \cdot T_e \cdot \Re\left( Z_f \cdot \Lambda(p_f, q_0) \right) + W^{ext}, \qquad (3.27)$$

where $Z_f$ and $p_f$ are constant parameters that depend on $Z_{1L}$, $Z_e$ and the noise temperatures of the probe $T_p$ and the preamplifier $T_e$. This demonstrates that the nuclear spin-noise spectrum is the real part of a pseudo-Lorentzian function. Accordingly, there exist solutions for which this function reduces to the pure in-phase Lorentzian, which corresponds to the SNTO condition.

As a consequence, in the SNTO condition, the observed resonance frequency can be different from the Larmor frequency (this corresponds to the case where $\Im(q_0) \neq 0$). The frequency offset, given by Equation (3.20), is $\frac{\Im(q_0)}{|q_0|^2} \omega_0$, and is *exactly* equal to the frequency shift given by Guéron's model [10]. Hence, the frequency shift of the chemical spectra corresponds exactly to the frequency offset seen on the spin-noise spectra. This result can be extended to the case of any noise-source temperatures. Optimal tuning condition corresponds to the simultaneous achievement of SNTO *and* FSTO. Mathematically, this is equivalent to:

$$FSTO = SNTO \Leftrightarrow \begin{cases} \Im(q_0) = 0, \\ \Im(Z_f p_f) = 0. \end{cases} \qquad (3.28)$$

Equation (3.28) describes the entire set of solutions of the SNTO problem, the general analysis of which is difficult. However, a sufficient (but not necessary) condition is given by:

$$\begin{cases} \Im(T^R) = 0, \\ \Im(Z_e) = 0. \end{cases} \Rightarrow SNTO = FSTO. \qquad (3.29)$$

In other words, if the impedance of the preamplifier and the four coefficients of the transmission matrix $T^R$ are real, then the SNTO condition is satisfied and the resonance frequency of the peak corresponds to the Larmor frequency (FSTO condition).

# IV. Features of the nuclear spin-noise spectra

## A. Dip or bump in spin-noise spectra at thermal equilibrium

Assuming a negligible noise contribution of the preamplifier to the spin dynamics ($T_e = 0$), equation (3.11) reveals several features of spin-noise line-shapes as a function of preamplifier impedances and transmission lengths in the SNTO condition. Firstly, the line width $(\lambda_2 + \lambda'_r)$ is affected by the impedance $Z_{3R}$ at the FSTO condition (equation (3.9)). Consequently it varies with the impedance of the preamplifier $Z_e$, and $\phi$. Secondly, the intensity of the spin-noise signal also depends on this impedance and can, at particular angles $\phi$, give different results from the model of McCoy and Ernst [4]. Indeed, the impedance $Z_{3R}$ can be small for particular phases of the transmission line since the measured impedances of the preamplifier are of the order of tens of Ω. If $G$ is large, i.e. $\lambda'_r \ll \lambda_r$, the circuit is designed in such a way that radiation damping is strongly reduced. The situation can even be such that $\lambda'^2_r + 2\lambda'_r\lambda_2 - \lambda^0_r\lambda_2$ becomes negative (for very large $G$, this corresponds to $\lambda^0_r > \lambda'_r$). Then instead of observing a decrease of noise at the Larmor frequency for a perfectly tuned probe (SNTO) at thermal equilibrium as predicted by the classical theory (equation (3.7)), an increase of noise, that is a bump and not the usual dip, is theoretically predicted and experimentally observed [29]. Conversely, if the impedance $Z_{3R}$ is large, $\lambda_r \simeq \lambda'_r$, the behavior becomes very similar to the predictions of McCoy and Ernst [4] or Guéron [10]. The transition from low impedance to high impedance can be easily obtained with a phase shift of 90°. Figure 2A illustrates this transition from a dip to a bump for three values of the transmission line phase $\phi$.

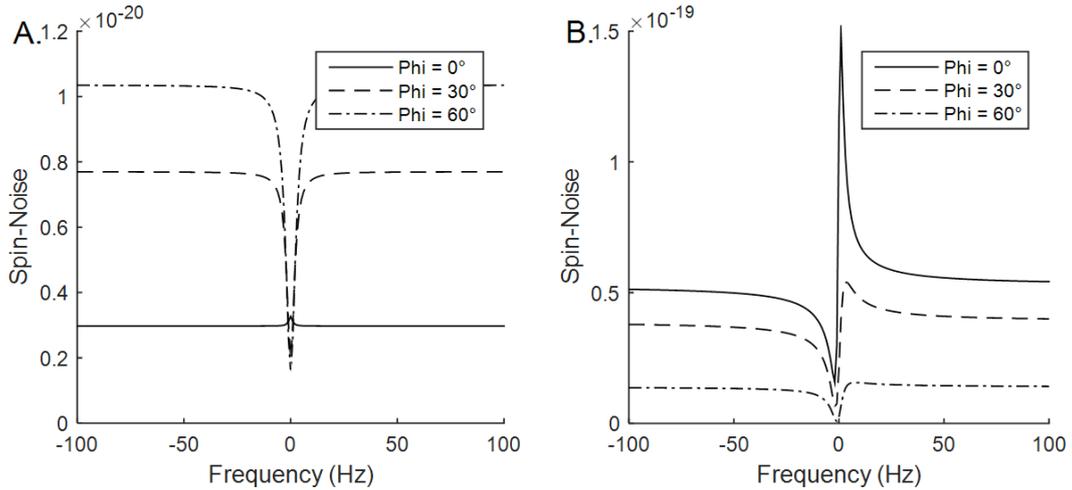

*Figure 2: Numerical simulation of spin-noise spectra for different $\phi$ parameters, under FSTO condition. On panel A, the noise temperature of the preamplifier is set to 0 K (no preamplifier noise) and Equation (3.12) was used. On panel B, the noise temperature of the preamplifier is set to 150 K and Equation (3.20) with a vanishing external noise ($W^{ext} = 0$). The different parameters were $\lambda_r = \lambda^0_r = 20\ Hz$, $\lambda_2 = 2\ Hz$.*

Figure 2A shows the nuclear spin-noise spectra using the no-noise preamplifier model described in equation (3.12) and assuming FSTO conditions. Perfect absorptive Lorentzian shapes are obtained with an average out-of-resonance noise level, linewidth and Larmor frequency noise level which depend on the transmission line. In particular, for $\phi = 0°$, a bump is predicted instead of the usual dip, in agreement with a previous report [29].

To explore a more realistic case, for which the effect of preamplifier noise was also considered, the calculation was carried out in the framework of the transmission matrices (Section III.C). We assumed $W^{ext} = 0$ and $T_e = 150\ K$, typical of low-noise preamplifiers. The result, given by equation (3.18), is shown in Figure 2B. The introduction of a noise source from the preamplifier induced different nuclear spin-noise shapes (not always pure in-phase Lorentzian) even if FSTO condition (Equation (3.9)) was assumed. Also, the average noise level was amplified by about one order of magnitude. The preamplifier contribution, even for low temperature noise, appears much higher than the probe contribution. Highly asymmetrical line shapes (for instance, for $\phi = 30°$) were observed. Moreover, the small bump seen on Figure 2A for $\phi = 0°$ was strongly enhanced with the appearance of a non-symmetrical high bump. The present simulations in FSTO conditions and taking into account the preamplifier noise which excites nuclear magnetization, lead to nuclear spin noise not always corresponding to SNTO conditions. This is the first numerical illustration that FSTO and SNTO conditions cannot be matched simultaneously. The comparison between the used assumptions for obtaining these two simulations (Figures 2A and 2B) indicates that different FSTO and SNTO conditions can be obtained only if noise produced by the preamplifier impedance can be fed back to the coil through the transmission line inducing excitations of the nuclear spins.

### B. *Experimental validation of the theoretical model*

In Figure 3, are reported 20 spin-noise spectra acquired for a series of $\psi$ values and a probe tuned and matched at 50 Ω using a vector network analyzer (PIMO conditions [30]). For $\psi = 10.1°$ and $\psi = 101°$, the nuclear spin-noise spectra appeared as almost perfect in-phase Lorentzian shapes, for the first one as a bump and the second one as a dip. On these two spectra the best least-squares fit to Equation (3.18) gave $\lambda_2 = 21.7\ Hz$, $\lambda_r = 148.2\ Hz$, $T_e = 175\ K$ and $W^{ext}$ which represented about 13% of the noise produced by the probe far from the NMR spin resonance. Finally, the value of $\phi_0$ was optimized, ($\phi_0 = -8.0°$), revealing that the second spectrum of Figure 3 corresponds to $\phi = 2.1°$.

In a next step, using the best fit parameters and the given $\phi$ values, for all panels the simulated curves obtained using Equation (3.18) were recomputed and superimposed (green lines) on the experimental measurements (red lines) (Figure 3). Good agreement is observed, which proves that spin-noise spectra are well described by Equation (3.18) and validates the theoretical model of Section III.C. This set of experiments shows that starting from the PIMO tuning condition it was possible to reach the SNTO condition by modifying the cable length between the preamplifier and the probe.

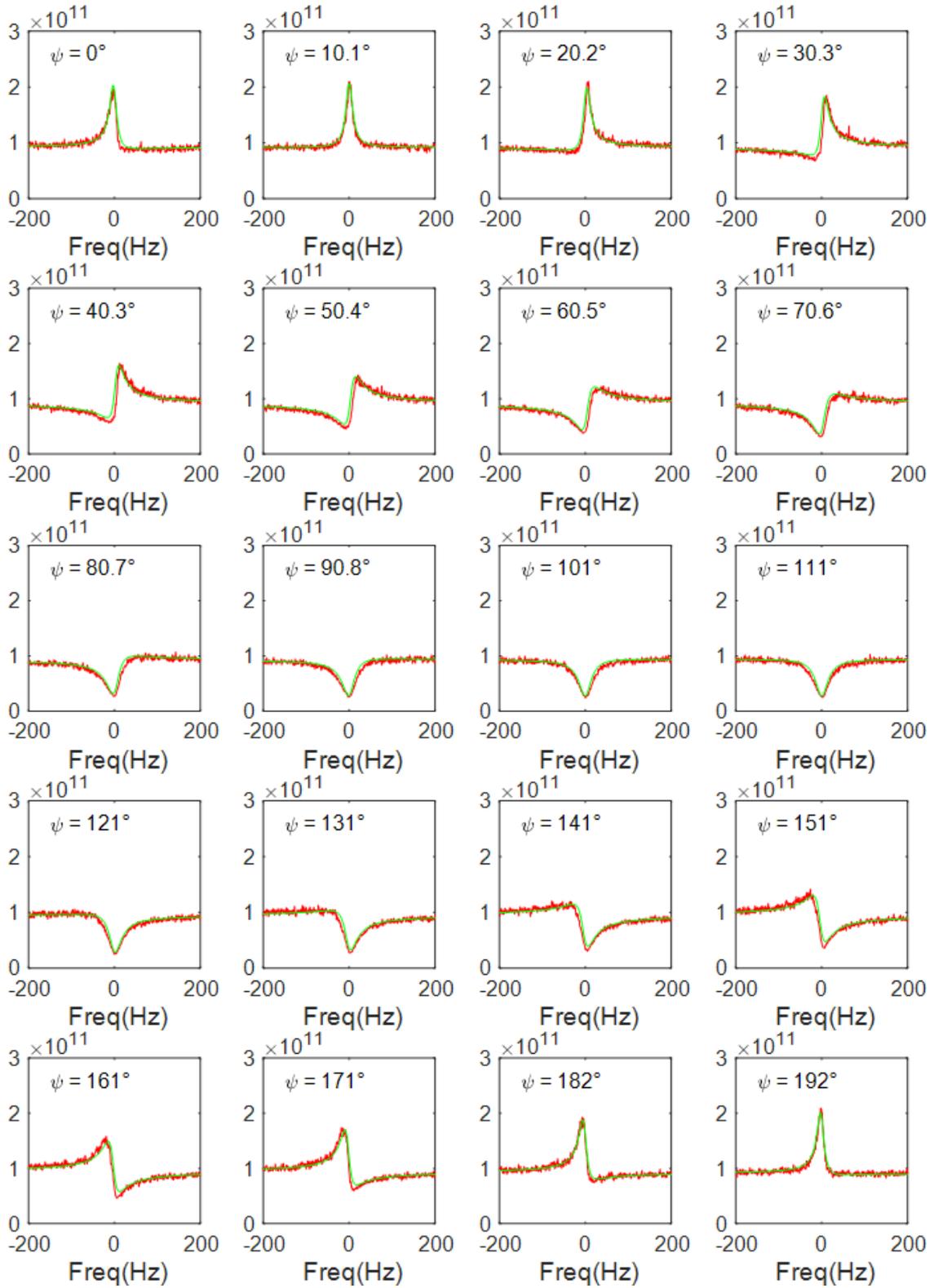

*Figure 3: Experimental nuclear spin-noise spectra of acetone acquired on a classical probe tuned at PIMO in red lines. The only differences between these 20 spectra were the variable transmission line phases $\psi$ which were incremented by step of 10.1°. SNTO conditions were found for $\psi = 10.1°$ with a bump signal and $\psi = 101°$ for which a dip was observed. Using the electronic component parameters, the four values $\lambda_r, \lambda_2, T_a, W^{ADC}$ were optimized on these two spectra. The best-fit curves computed using Equation (3.18) were superimposed. Then for all the others experimental spectra, the simulated spectra were recomputed by only changing the phase values $\psi$ and keeping all the other parameters constant. The simulated curves are also superimposed in green lines. A good agreement can be observed.*

## C. Obtaining the same FSTO and SNTO conditions

The last issue to address is how to practically ensure simultaneous FSTO and SNTO conditions. Solving equation (3.28) was impossible in the general case, we have therefore chosen to address the sufficient condition (3.29). We verified experimentally that the PIMO condition almost satisfied equation (3.29) for two different phases $\phi$. As a result, the conditions PIMO and FSTO entailed the condition SNTO.

Starting from the PIMO condition, we explored the variation of peak resonance frequencies on spectra acquired after small-excitation pulses for $\psi$ ranging from 0 to 190° in 10° steps. It can be shown that this variation is described by a sinusoid [35]. The mean shift corresponded to the Larmor frequency and allowed the determination of phases $\psi$ for which the FSTO condition was fulfilled. An illustration of this dependence is reported in Fig. 4A. The best-fit to a sinusoid of the experimental measurements allowed the determination of the FSTO conditions which were fulfilled for $\psi$ = 64.5° and 154.5°.

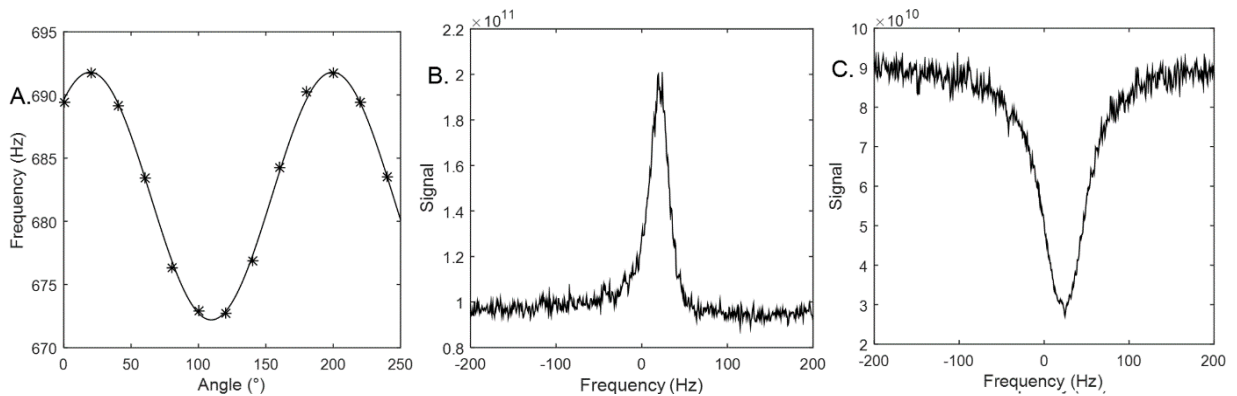

*Figure 4: Panel A: Variation, as a function of the transmission line phase $\psi$, of the nuclear resonance frequency of acetone measured on spectra acquired after small excitation pulses, for a probe tuned according to the PIMO condition. Asterisks show the experimental measurements and the solid line represents the best-fit sinusoid curve. From this analysis the FSTO conditions could be determined, they corresponded to $\psi = 64.5°$ and $\psi = 154.5°$. The two nuclear spin-noise spectra acquired with these phases are shown in panels B and C, respectively. Their shapes were almost perfect in-phase Lorentzian with a bump and a dip, respectively and a linewidth also affected by the transmission line phase.*

Finally, for these two phases, spin-noise spectra were acquired and the results are reported in Figure 4B and C. They corresponded to almost perfect in-phase Lorentzian shapes with a dip for $\psi = 154.5°$ and a bump for $\psi = 64.5°$. As a consequence they correspond to the SNTO condition and proved that starting from PIMO condition and by monitoring the frequency shift variation as a function of the angle $\phi$ it was possible to simultaneously ensure FSTO and SNTO.

Careful shape analysis nevertheless reveals that the SNTO condition was not perfectly achieved for $\psi = 64.5°$. The peak is slightly asymmetric. We have therefore analyzed the reliability of the FSTO=SNTO condition and its stability, using the numerical simulation approach based on transmission matrices (Section III.C). Starting from PIMO condition and changing the tuning or the matching capacitance by 0.5% induced the disappearance of the capability of obtaining same FSTO and SNTO conditions. In the case of probe and sample cooled down at $T_p = -60°C$, a difference between FSTO and SNTO has been experimentally shown [28]. We assumed that the only effect of this cooling was the reduction of the resistance $R_c$ by a factor 2. The threshold on the tuning capacitance for obtaining same PIMO, FSTO, and SNTO conditions was then reduced from 0.5% to 0.1%. Finally, the threshold for cold probe with a $Q$ factor of 8000 and thus a resistance of $R_c = 0.01\ \Omega$, was found to be 0.02%, a value typically below of what could be achieved based on mechanical and electronic precisions. This result illustrates that even if in theory it should be possible to match different tuning conditions (PIMO, FSTO, SNTO) even on a cold probe, it can be expected to be difficult to achieve on an experimental point of view.

For the two optimal configurations for which FSTO and SNTO are simultaneously fulfilled, either a bump or a dip with an in-phase Lorentzian line shape and no frequency shift are observed. For these spectra, the signal-to-noise ratio can be defined as:

$$SNR = \frac{|S_0 - S_{<>}|}{\sigma_{S_0}} \tag{4.1}$$

Where $S_{<>}$ is the average noise level out-of-resonance and $S_0$ and $\sigma_{S_0}$ are the noise level at nuclear resonance and its associated standard deviation which scales as the square-root of the number $n$ of scans: $\sigma_{S_0} \propto S_0\sqrt{n}$. Denoting $k = S_0/S_{<>}$, the signal-to-noise ratio scales as $|1 - 1/k|\sqrt{n}$ [8, 21]. For a bump, this SNR is limited to $\sqrt{n}$, while for a dip it can reach larger value if $k < 0.5$. Such a condition can experimentally be obtained by reducing the noise contribution of the preamplifier and increasing the radiation damping rate. These last two conditions are fulfilled for the dip configuration. As a whole, the dip configuration appears, in general, as the best one for observing nuclear spin noise.

## V. Discussion

### A. Comparison with the McCoy and Ernst theory

In Section III.D, we have shown that in a restricted range of frequency around the Larmor frequency, the spin-noise spectrum can be represented by $a\,\Re\left(\Lambda[p,q]\left(\frac{\omega-\omega_0}{\omega_0}\right)\right)$ where $a$ is a real coefficient and $p$ and $q$ are complex values. This result proves that in general the shape of the spin-noise spectrum can be represented as a mixture of absorptive and dispersive Lorentzian functions.

This is also the case of the classical equation (3.7) derived by McCoy and Ernst [4] and Guéron [10], this explains why it was generally impossible by a simple fitting procedure to evidence the weaknesses of this usual model. Indeed, except for the particular cases where bumps at thermal equilibrium and using classical probe are observed [29], experimental spin-noise spectra can always be described by Equation (3.7) through adjusting $\lambda_r$, $\lambda_2$, $\omega_0$ and constant noise levels for the coil and the preamplifier ($W_c$). The alternative way to reveal the weakness of this theory was by showing that for SNTO condition the frequency shift can experimentally be different from zero [28]. In our model, the observation of a ``bump'' and non-vanishing frequency shift for SNTO are now explained, since a pseudo-Lorentzian function with a complex-valued parameter $q$ can be transformed into a pseudo-Lorentzian function with real-valued parameter $q$ through a translation in the frequency domain (equation (3.20)), that is with a frequency shift contribution. We consequently have a mathematical definition of the SNTO condition which is not dependent on matching the resonance frequency of the electronic circuit and the Larmor frequency, the latter corresponds to the physical definition of FSTO condition. In that sense, the contradiction between experiments and theoretical derivations put forward in reference [28] is definitively solved.

## B. *A phenomenological equation*

For practical applications of NMR spectroscopy, in particular using a cold probe, it is desirable to have a simplified model reproducing the essential characteristics described by the general one presented here. The usual inability to mathematically distinguish the general model from the McCoy and Ernst's model provides clues for obtaining a phenomenological equation which could be used to fit the experimental measurements and to obtain relevant parameters (transverse relaxation and radiation damping rates, Larmor resonance frequencies, offset in tuning). If one considers a spectrum acquired at SNTO, the FSTO condition is not automatically fulfilled. According to Equation (3.20), the term $\delta\omega = \omega - \omega_0$ in Equation (3.8) has to be replaced by $\delta\omega' = \delta\omega - \zeta$, where $\zeta$ is the frequency shift dependent on the difference between the Larmor frequency and the frequency corresponding to the FSTO condition, $\omega_{\text{FSTO}}$. The second correction consists in noting that the SNTO condition does not require a constant noise level out-of-resonance. Since the frequency ranges of NMR spectra are much narrower than the bandwidths of detection circuit, a linear dependence seems sufficient to represent this variation of the average noise level. This finally leads to the following phenomenological equation:

$$W_t = A \cdot \frac{1 + \lambda_r^0 \, a(\delta\omega')}{[1 + \lambda_r' \, a(\delta\omega')]^2 + [\Delta + \lambda_r' \, d(\delta\omega')]^2} + B\delta\omega + C. \tag{5.1}$$

The parameter Δ represents the tuning-dependent frequency offset between the actual resonance frequency for spin-noise and the perfect SNTO condition leading to a phase-mixed Lorentzian resonance shape [28].

### C. Equivalent electronic circuit

The theoretical derivation also physically and mathematically defines what is meant by the FSTO condition. It corresponds to an equivalent impedance, $Z_{6R}$, at the reference plan #6, which is purely real. In terms of radiation damping physics this is equivalent to a current in phase with the source voltage produced by the precessing magnetization [9], inducing a feedback field in quadrature to the precessing magnetization. Mathematically the FSTO condition corresponds to Equation (3.9) which appears to be dependent on the tuning capacitance but also on all the other components of the electronic reception circuit. It is particularly dependent on the length of the transmission line through the phase $\phi$. This is in agreement with previous experimental observations which have shown the existence of a large number of tuning and matching conditions for obtaining SNTO and have revealed the dependence on the transmission line of the SNTO conditions [28, 29].

Finally, the general demonstration and calculation carried out here, prove that the spin-dynamics interaction between the magnetization and the electronic circuit can be modeled by a simple equivalent RLC circuit [9]. Obviously, the effective quality factor is not given by the quality factor of the coil $Q = L\omega/R_c$ but depends on all electronic components of the reception circuit, explaining the appearance of the modified radiation damping rate $\lambda'_r$ in Equations (3.10) and (5.1) and the introduction of an apparent quality factor $Q' = Q / \left(1 + \frac{G}{R_c X}\right)$, also introduced experimentally for explaining the difference between the extracted $Q$ parameters and the ones claimed by the probe manufacturers [8, 28, 29, 36].

## VI. Conclusion

Nuclear spin noise in NMR has been observed for the first time more than 25 years ago and described by McCoy and Ernst's equation [4] which was derived assuming that all properties of the electronic circuit leading to radiation damping can be reproduced by an equivalent RLC circuit. Recently, it was experimentally shown that the predictions of this model are incorrect since pure in-phase spin-noise spectra can be obtained with a non-vanishing frequency shift, i.e. a radiation damping field not in quadrature to the magnetization [28]. The model developed in the present article solves this difficulty by introducing a careful description of the detection circuit. The discrepancy appears to result from magnetization excitations due to the fluctuating current within the finite impedance of the preamplifier coupled to the coil through a transmission line which has the incorrect length. The quality of this model was assessed by extensive comparisons between simulated and experimental spectra performed for a large series of transmission line phases $\phi$. We also theoretically predict experimental spin-noise spectra with pure in-phase Lorentzian shape and a bump rather than the usual dip in the average noise level. The latter solution appears as the best one in terms of signal-to-noise ratio per time unit for spin-noise detection. We also show the influence of the preamplifier impedance and transmission line phase on the observed resonance linewidth. Finally, the model reveals that experimental conditions exist for which the FSTO and SNTO conditions match but are almost unattainable with a cold probe due to uncertainties in the determination of the tuning and matching capacitor values.

Calculation of spin-noise spectra with the present model requires the knowledge of all electronic components of the detection circuit, a feature which is feasible on a room temperature probe but which is usually beyond what an NMR spectroscopist can do on a commercial cold probe. As a consequence, we have introduced a phenomenological equation valid for one spin-species which allows one to best-fit experimental measurements in order to have access to NMR parameters (effective radiation damping rate, relaxation rate, and resonance frequency).

## *Acknowledgments*

This research was supported the Agence Nationale de Recherche (project IMAGINE, ANR project 12-IS04-0006).

# Nuclear spin noise in NMR, revisited
# Supplemental material

Guillaume Ferrand,[1] Gaspard Huber,[2] Michel Luong,[1] Hervé Desvaux[2,a]

1. Laboratoire d'ingénierie des systèmes accélérateurs et des hyperfréquences, SACM, CEA, Université Paris-Saclay, CEA/Saclay, F-91191 Gif sur Yvette, France.

2. Laboratoire Structure et Dynamique par Résonance Magnétique, NIMBE, CEA, CNRS, Université Paris-Saclay, CEA/Saclay, 91191 Gif sur Yvette, France.

a) herve.desvaux@cea.fr

## I. Supplementary elements for the theoretical developments

### A. Demonstration of Equation (3.15)

In the general case, if $T$ is a transmission matrix transforming the voltage and intensity $U_1^+, i_1^+$ into $U_2^+, i_2^+$, i.e. $\begin{pmatrix} U_2^+ \\ i_2^+ \end{pmatrix} = \begin{pmatrix} T_{11} & T_{12} \\ T_{21} & T_{22} \end{pmatrix} \begin{pmatrix} U_1^+ \\ i_1^+ \end{pmatrix}$, for the inverse transmission matrix, by convention, the signs of the intensities have to be inverted ($i_1^- = -i_1^+$). Since for here-considered electronic components the system is symmetrical, the inverse transmission matrix is:

$$\begin{pmatrix} U_1^- \\ i_1^- \end{pmatrix} = \begin{pmatrix} T_{22} & T_{12} \\ T_{21} & T_{11} \end{pmatrix} \begin{pmatrix} U_2^- \\ i_2^- \end{pmatrix}. \quad (A.1)$$

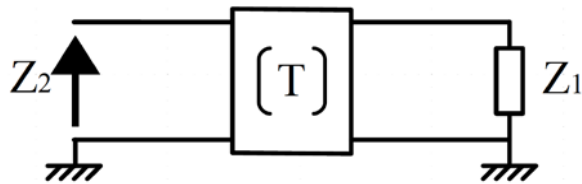

*Figure S.1 Schematic of a transmission matrix transforming $Z_1$ to $Z_2$.*

Considering the scheme of Figure S.1, from the impedance $Z_1$ and the transmission matrix $T$ the impedance $Z_2$ can be computed:

$$Z_2 = \frac{U_2}{i_2} = \frac{Z_1 T_{11} + T_{12}}{Z_1 T_{21} + T_{22}}. \quad (A.2)$$

Conversely, for the transformation of $Z_2$ into $Z_1$, one obtains:

$$Z_1 = \frac{Z_2 T_{22} + T_{12}}{Z_2 T_{21} + T_{11}}. \quad (A.3)$$

According to the sign convention of Figure 1, the ratio $U_{sn}/i_{sn}$ is equal to $-Z_{6L}$, which can be computed thanks to Equation A.3. The result can be applied to Equation 3.14, replacing $i_{sn}$ by $U_{sn}/-Z_{6L}$. Finally, $U_z$ is obtained:

$$U_z = \frac{Z_e U_{sn}}{Z_e T_{22}^{sn} + T_{12}^{sn}}. \quad (A.4)$$

This was used to obtain the spectral density (Equation 3.15).

Preamplifiers are characterized by noise factors $F_{dB}$, expressed in dB, which describes the signal-to-noise ratio degradation between the input and the output. The relation between $F_{dB}$ and the equivalent temperature $T_e$ used in the article (Equation 3.16 and numerical applications, is:

$$T_e = T_0 (10^{F_{dB}/10} - 1), \quad (A.5)$$

with $T_0$ a reference temperature equal to 290 K.

## B. Demonstration of the pseudo-Lorentzian shape (Equation 3.27)

There are two useful properties of pseudo-Lorentzian functions $\Lambda$, defined by Equation 3.19:

$$\frac{1}{a \cdot \Lambda[p,q] + b} = \frac{1}{a+b} \cdot \Lambda\left[-p \cdot \frac{a}{b + a(1+p)}, q \cdot \frac{b+a}{b + a(1+p)}\right], \quad (B.1)$$

And:

$$a \cdot \Lambda[p,q] + b = (a+b) \cdot \Lambda\left[p \cdot \frac{a}{a+b}, q\right]. \quad (B.2)$$

Equation 3.23 of the main text can be written as:

$$Z_{1L} = \frac{Z_{rs} T_{11}^R + T_{12}^R}{Z_{rs} T_{21}^R + T_{22}^R} = \frac{T_{11}^R}{T_{21}^R} + \left(T_{12}^R - \frac{T_{22}^R T_{11}^R}{T_{21}^R}\right) \cdot \frac{1}{Z_{rs} T_{21}^R + T_{22}^R} \quad (B.3)$$

where $Z_{rs}$ is a pseudo-Lorentzian function as shown in the main text. By using successively the two relations B.1 and B.2, it becomes clear that $Z_{1L}$ is a pseudo-Lorentzian function.

Starting from Equation 3.26, and defining $Z_{1f} = \frac{Z_e Z_{1L}}{Z_e + Z_{1L}} = Z_e - \frac{Z_e^2}{Z_e + Z_{1L}}$ and $K_{2f} = \frac{\frac{T_0}{T_e} Z_e + Z_{1L}}{Z_e + Z_{1L}} = 1 + \frac{\left(\frac{T_0}{T_e} - 1\right) \cdot Z_e}{Z_e + Z_{1L}}$, we can show using Equations B.1 and B.2 that these two terms, $Z_{1f}$ and $K_{2f}$ are pseudo-Lorentzian functions since $Z_{1L}$ is a pseudo-Lorentzian function. According to Equation B.1, they have the same $q$ parameter which we denote $q_0$ since they have the same denominator. We define $p_{f1}$ and $p_{f2}$ as the $p$ parameters of $Z_{1f}$ and $K_{2f}$, respectively and $Z_{1\infty}$ and $K_{2\infty}$, the values of $Z_{1f}$ and $K_{2f}$ for far off-resonance condition ($\omega \to \infty$). Then, the product $Z_{1f}K_{2f}^\star$ can be written as:

$$Z_{1f}K_{2f}^\star = Z_{1\infty}K_{2\infty}^\star \left(1 + \frac{p_{f1}}{1+jq_0x} + \frac{p_{f2}^\star}{1-jq_0x} + \frac{p_{f1}p_{f2}^\star}{1+q_0^2x^2}\right). \quad (B.4)$$

First, if $q_0$ is real-valued, we have:

$$\begin{cases} \Re\left(\frac{p_{f2}^\star Z_{1\infty}K_{2\infty}^\star}{1-jq_0x}\right) = \Re\left(\frac{p_{f2}Z_{1\infty}^\star K_{2\infty}}{1+jq_0x}\right), \\ \Re\left(\frac{Z_{1\infty}K_{2\infty}^\star p_{f1}p_{f2}^\star}{1+q_0^2x^2}\right) = \Re(Z_{1\infty}K_{2\infty}^\star p_{f1}p_{f2}^\star)\Re\left(\frac{1}{1+jq_0x}\right). \end{cases} \quad (B.5)$$

According to Equations B.4 and B.5:

$$\begin{cases} \text{if: } Z_{eq} = Z_{1\infty}K_{2\infty}^\star + \dfrac{p_{f1}Z_{1\infty}K_{2\infty}^\star + p_{f2}Z_{1\infty}^\star K_{2\infty} + \Re(p_{f1}p_{f2}^\star Z_{1\infty}K_{2\infty}^\star)}{1+jq_0x}, \\ \text{then: } \Re(Z_{eq}) = \Re(K_{1f}Z_{2f}^\star). \end{cases} \quad (B.6)$$

It demonstrates that $W_{tot}$ in Equation 3.26 can be written as the real part of a pseudo-lorentzian function, $Z_{eq}$.

Now, if $q_0$ is not real-valued ($\Im(q_0) \neq 0$), by using Equation 3.20 the transformation of $x$ into $x' = x - \frac{\Im(q_0)}{|q_0|^2}$ allows the restoration of real $q_0' = |q_0|^2/\Re(q_0)$ parameter. Noting that $\Im(q_0) = \Im(-q_0^\star)$, thus Equation B.4 remains valid, after the transformation of $x, q_0, p_{f1}$ and $p_{f2}$ into a set $x', q_0', p_{f1}'$ and $p_{f2}'$ according to Equation 3.20. Equations B.5 and B.6 must be modified with these new parameters. Finally, Equation 3.20 can be used in Equation B.6 to transform $q_0'x'$ into $\frac{|q_0|^2}{\Re(q_0)}\left(x - \frac{\Im(q_0)}{|q_0|^2}\right)$. This demonstrates the pseudo-lorentzian shape of the spin-noise resonance *whatever* the value of the $q_0$ parameter and whatever the noise temperature of the preamplifier.

## C. McCoy and Ernst's equation and the pseudo-Lorentzian function

Equation 3.6 introduced by McCoy and Ernst was used for several decades for describing spin-noise spectra.[1] This equation can indeed be written as a pseudo-Lorentzian function. In order to prove it, let $Z_c$ and $Z_l$ be defined as: $Z_c = R_c + jL\omega - j(C_t\omega)^{-1}$ and $Z_l = R_c\lambda_r\big(a(\delta\omega) + jd(\delta\omega)\big)$. With these definitions, $W_t$ of Equation 3.6 can be written as:

$$W_t = \frac{2}{\pi} \cdot \frac{k_B T_p}{\omega^2 C_t^2} \cdot \frac{\Re(Z_c) + \frac{\lambda_r^0}{\lambda_r}\Re(Z_l)}{|Z_c + Z_l|^2}. \tag{C.1}$$

Similar expression is obtained for Equation 3.8 (model with non-infinite input impedance of the preamplifier). Equation C.1 can be expressed in a similar way as Equation 3.25 and thus the same transformation can be used. Indeed:

$$W_t = \frac{2}{\pi} \cdot \frac{k_B T_p}{\omega^2 C_t^2 R_c^2} \cdot \Re\left(\frac{R_c^2}{Z_c + Z_l} \cdot \left(\frac{\frac{\lambda_r^0}{\lambda_r}Z_l + Z_c}{Z_c + Z_l}\right)^\star\right). \tag{C.2}$$

Equation C.2 can be written as:

$$W_t = \frac{2}{\pi} \cdot \frac{k_B T_p}{\omega^2 C_t^2 R_c^2} \cdot \Re\left(\frac{R_c^2}{Z_c + Z_l} \cdot \left(\frac{\lambda_r^0}{\lambda_r}\left(1 - \frac{\left(1 - \frac{\lambda_r^0}{\lambda_r}\right)Z_c}{Z_c + Z_l}\right)\right)^\star\right). \tag{C.3}$$

Since $Z_l = \frac{R_c \lambda_r}{\lambda_2 - j(\omega - \omega_0)}$, $Z_c + Z_l$ is a pseudo-Lorentzian function defined by:

$$Z_c + Z_l = (R_c + j(L\omega - (C_t\omega)^{-1}))\Lambda\left[\frac{R_c}{R_c + jL\omega - j(C_t\omega)^{-1}} \cdot \frac{\lambda_r}{\lambda_2}, -\frac{\omega_0}{\lambda_2}\right]\left(\frac{\omega - \omega_0}{\omega_0}\right). \tag{C.4}$$

According to Equation B.1, with $b = 0$, $Z_{1f} = \frac{R_c^2}{Z_c + Z_l}$ is a pseudo-Lorentzian function with parameters:

$$\begin{cases} p_{1f} = \dfrac{-R_c}{\left(\frac{\lambda_2}{\lambda_r} + 1\right)R_c + \frac{\lambda_2}{\lambda_r}(jL\omega - j(C_t\omega)^{-1})}, \\ q_0 = -\dfrac{R_c + jL\omega - j(C_t\omega)^{-1}}{\left(\frac{\lambda_2}{\lambda_r} + 1\right)R_c + \frac{\lambda_2}{\lambda_r}(jL\omega - j(C_t\omega)^{-1})} \cdot \dfrac{\omega_0}{\lambda_r}. \end{cases} \tag{C.5}$$

In the same way, $K_{2f} = \frac{\lambda_0}{\lambda_r}\left(1 - \frac{\left(1 - \frac{\lambda_r^0}{\lambda_r}\right)Z_c}{Z_c + Z_l}\right)$ is a pseudo-Lorentzian function. Especially, if $\lambda_r^0 = \lambda_r$, $K_{2f} = 0$, and the nuclear spin-noise shape corresponds to the shape of $Z_{1f}$ given by Equation C.5. Otherwise, the development of the Equations is similar to Equation B.6. Whatever the value of $\lambda_r^0$, the final quality factor is given by $q_0$ (Equation C.5).

If $(L\omega - (C_t\omega)^{-1}) = 0$, the coefficients $q_0$, $p_{1f}$ and $p_{2f}$ are purely real. The quality factor $q_0$ is given by: $q_0 = -\frac{\omega_0}{\lambda_2 + \lambda_r}$. This corresponds to the quality factor of Equation 3.7. This corresponds to the simultaneous SNTO and FSTO condition. Otherwise, $p_{1f}$, $p_{2f}$ and $q_0$ are complex-valued, and there is a frequency shift contribution equal to $\zeta = \frac{\Im(q_0)}{|q_0|}$ but simultaneously the shape of the spin-noise resonance is not that of an in-phase Lorentzian function.

### D. Demonstration of the sinusoidal shape of Figure 4A

The transmission matrix of a transmission line is given by $T_L = \begin{bmatrix} \cos\phi & jZ_0 \sin\phi \\ \frac{j\sin\phi}{Z_0} & \cos\phi \end{bmatrix}$.

In the condition of Equation 3.29 (that is the solution proposed for simultaneous SNTO and FSTO condition), if a phase shift is added to $T^R$, between the probe and the preamplifier, the total transmission line is given by the matrix product $T_L \cdot T^R$. Using Equations B.1 and B.2 for the transmission matrix product $T_L \cdot T^R$, where $T^R$ is real-valued, we observe that the resulting parameter $\frac{\Im(q)}{|q|^2}$, and thus the frequency shift, is a function of $\sin 2\phi$. If the impedance of the preamplifier is purely real (as defined in Equation 3.29), then, the final frequency shift $\frac{\Im(q_0)}{|q_0|^2}$ also varies as $\sin 2\phi$. This demonstrates the sinusoidal shape of the frequency shift, for the specific case of Equation 3.29.

Especially, in the PIMO condition, there is a phase $\phi^0$ for which the transmission matrix $T^R$ verifies Equation 3.29. In other words, there is a transmission line matrix $T_L^0$ for which $\Im(T_L^0 \cdot T^P) = 0$, where $T^P$ is the transmission matrix of the probe in PIMO conditions. Consequently, adding a transmission line $T_L$ to a probe matched by PIMO is equivalent to adding a transmission line described by $T_L \cdot T_L^{0^{-1}}$ to a probe with transmission matrix $T^R$ that verifies Equation 3.29. Consequently, $\frac{\Im(q_0)}{|q_0|^2}$ varies according to $\sin(2\phi - 2\phi^0)$, where $\phi^0$ is the phase parameter of $T_L^0$. This validates the shape of Figure 4A. For other conditions (especially CTO conditions), we experimentally observed that the shape of $\frac{\Im(q_0)}{|q_0|^2}$ is not a sinusoidal function of $\phi$ (see below).

## II. Dependence of the spin-noise spectra on the tuning parameters

As another procedure for exploring the predictions of the theoretical developments, we report here an experimental study for which the probe was not tuned at PIMO but in contrast, the tuning and matching capacitances were adjusted for each phase of the transmission line for ensuring SNTO condition, in a procedure reminiscent to previous studies.[2-4]

NMR experiments have been performed on a Bruker Avance DRX500 spectrometer equipped with an inverse broad band probehead with z-gradient, operating at 500 MHz for proton. The sample was made of 90% methyl-isopropyl ketone and 10% deuterated benzene for magnetic field lock purpose.

Transmission phase were modified by changing the value of the phase shifter angle, placed between the preamplifier and the probe head. This device has been previously calibrated by connection to a transceiver. For each phase shift, the tuning $C_t$ and matching $C_m$ capacitors were adjusted in CTO condition (emission mode). Then keeping the value of $C_m$ constant, the tuning capacitor $C_t$ was adjusted for ensuring SNTO condition (reception mode). For phases between 10 and 35°, it was not possible to adjust $C_t$, because of a limited accessible range. Also for $\psi = 0°$ the tuning was achievable with two extreme values of $C_t$, so the parameter values corresponding to the two measurements are reported.

After the achievement of this tuning procedure, a classical proton NMR spectrum obtained after a small flip-angle excitation pulse (ca. 7°) and a spin-noise spectrum composed of 8 FIDs, each one containing 512k points acquired in 52.4 s were recorded. From these two spectra, four NMR parameters were extracted: the resonance frequency and thus the frequency shift and the linewidth of $^{12}C\underline{H}_3CO$ signal from the classical FID, noise level out-of and at resonance from the noise spectra. We, indeed, preferred to avoid directly exploiting spin-noise spectra for extracting frequency shifts and linewidths since small offsets relative to perfect SNTO conditions would lead to significant errors in the determination of these two parameters.[4] Examples of these spectra acquired after a small flip-angle excitation pulse and using the nuclear spin-noise scheme are reported in Figure S.2.

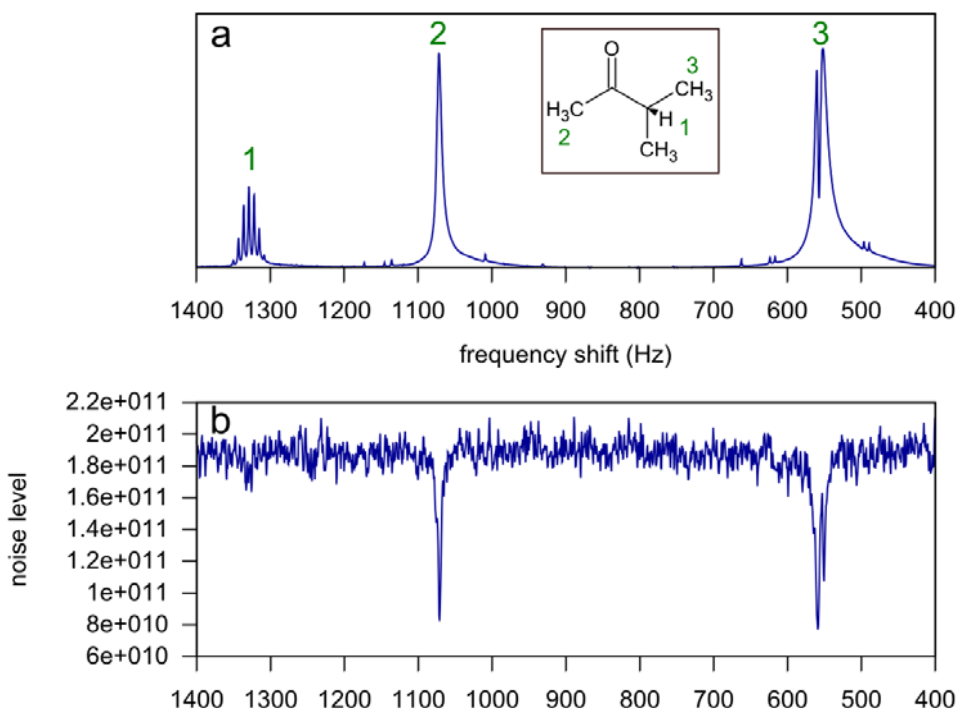

*Figure S.2. Examples of aliphatic parts of $^1H$ spectra acquired with a small flip-angle excitation pulse (A) on methyl isopropyl ketone (in insert) and with nuclear spin-noise scheme (B). These spectra have been acquired with a phase $\psi=0°$ of the phase shifter. The linewidth and resonance frequency of signal 2 measured on spectrum A were used in the following steps of analysis and the average noise level out-of-resonance (1.90 $10^{11}$) and at resonance (about 7.7 $10^{10}$) measured on the spin-*

*noise spectrum (B). A significant frequency shift contribution can easily be detected through the shape of the doublet 3 (see Reference [5]).*

Figure S.3 illustrates how the nuclear spin-noise spectra are affected by the transmission line phase $\psi$, that is, according to the theoretical developments (Equations 3.10, 3.11 and 3.12) the apparent radiation damping rate $\lambda'_r$. The linewidths, the depths of the resonance dip as well as the average noise levels out-of-resonance were affected by the change of this parameter and the induced changed of tuning $C_t$ and matching $C_m$ capacitors needed for ensuring the observation of SNTO condition.

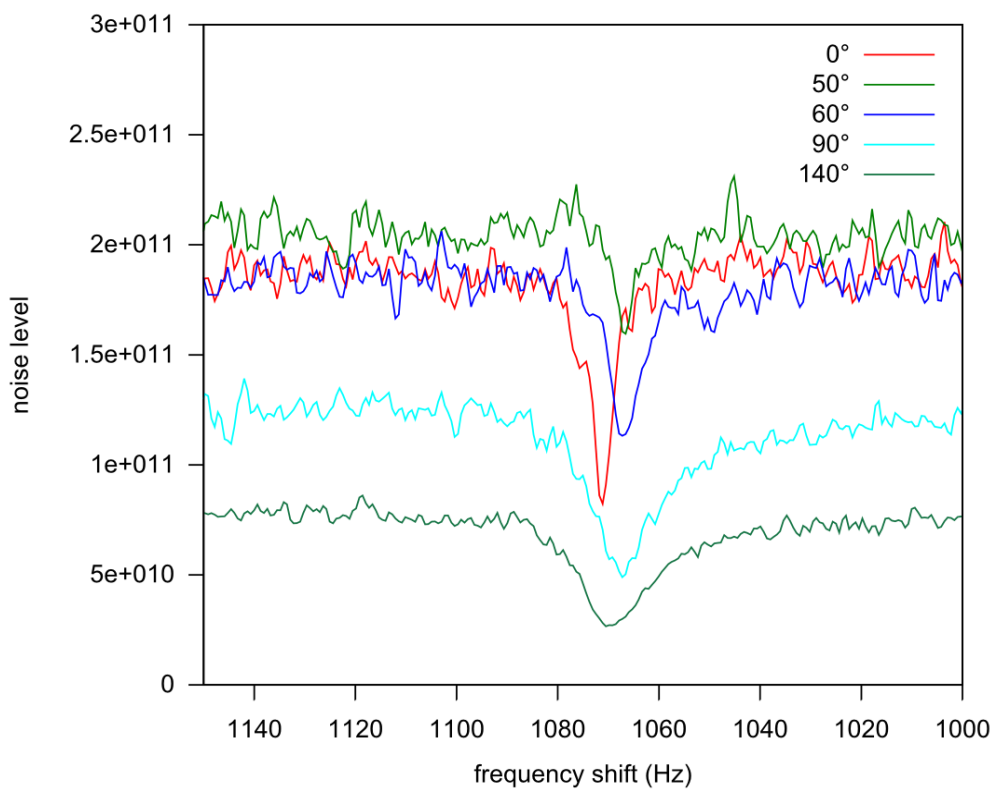

*Figure S.3. $^1$H nuclear spin-noise sub-spectra corresponding to CH$_3$CO resonance of methyl-isopropyl ketone as a function of the transmission phase $\psi$.*

In Figure S.4, we report the variation of these four parameters (linewidths, resonance frequencies, noise level at resonance and noise level out-of-resonance) for the whole series of explored transmission phases $\psi$. Firstly one can notice a significant variation of the difference of resonance linewidths as a function of phase $\psi$; the differences were defined by subtracting the full linewidth at half height of $^{12}$CH$_3$CO resonances to that of $^{13}$CH$_3$CO resonances. In such a way, variation of field homogeneity was circumvented; also most of the natural linewidths ($\lambda_2$ contribution) were cancelled. The apparent radiation damping contributions ($\lambda'_r = \pi$ FWMH) vary from about 0 to 51 Hz, clearly illustrating how the conjunction of the input impedance of the preamplifier and the transmission phase $\psi$ affects these contributions, as predicted by Equation 3.10. In comparison a more restricted variation of the resonance frequencies corresponding to a maximum frequency shift contribution of 19 Hz as a function of the phase shifter angle $\psi$ was detected; it illustrates that with the present electronic system, the SNTO and FSTO conditions can be different but with an extent much more restricted than that observed with a cold probe.[4] For the noise levels, when the radiation damping contributions are at smallest ($\psi \sim 40°$) the average noise level out-of-resonance as well as the noise level at resonance were at maximum. Since uncertainties scale as the amplitudes of the signals in spin-noise experiments,[6] it indicates that this configuration ($\psi \sim 40°$) is also the less favorable for spin-noise determination of NMR parameters. Finally, one can notice that for the four explored parameters their variations as a function of the angle $\psi$ are not a sinusoidal function as discussed in Section I.D of the Supplemental Material.

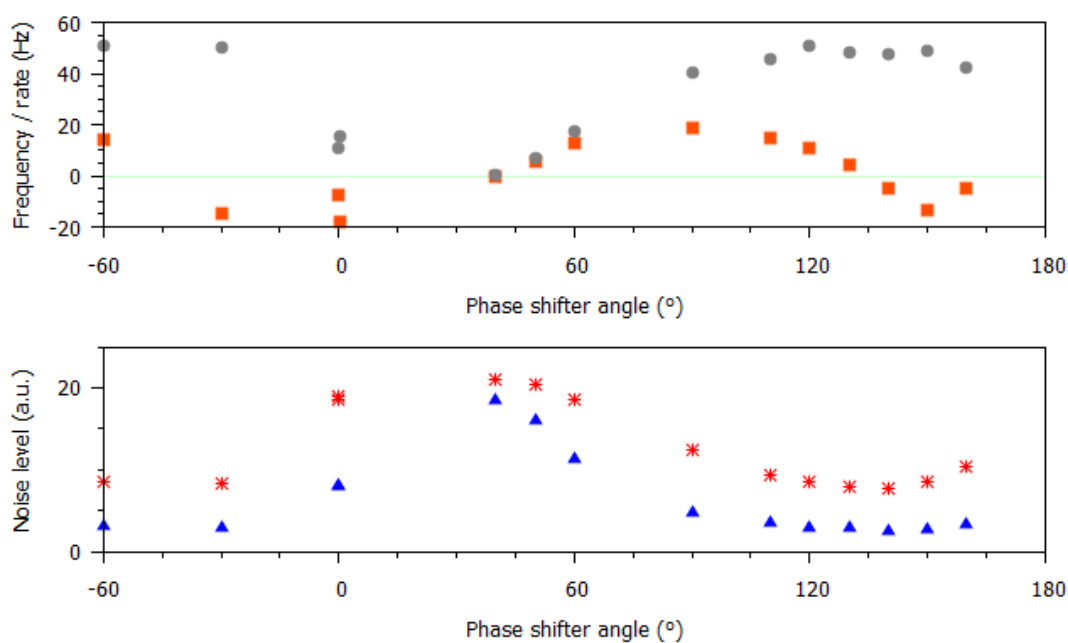

*Figure S.4. Parameters deduced from (a) excitation pulse and (b) spin-noise spectra as a function of the transmission phase $\psi$. (a) Relative variations of the resonance frequencies of $CH_3CO$ resonance of methyl-isopropyl ketone referred to that in the FSTO conditions (squares, the absolute values are directly given as the frequency shift contribution), and of the linewidths (circles, the values are directly given as the radiation damping rate $\lambda'_T$). (b) Variation of the noise levels in arbitrary unit, at resonance (triangles) and out-of-resonance (stars) for the different phase values $\psi$.*